# Orientation dependence of the nano-indentation behaviour of pure tungsten

Hongbing Yu[1*], Suchandrima Das[1], Haiyang Yu[2], Phani Karamched[2], Edmund Tarleton[1], Felix Hofmann[1†]

1) Department of Engineering Science, University of Oxford, Parks Road, Oxford, OX1 3PJ, UK

2) Department of Materials, University of Oxford, Parks Road, Oxford, OX1 3PH, UK

[*] *hongbing.yu@eng.ox.ac.uk*        [†] *felix.hofmann@eng.ox.ac.uk*

**Abstract**: Coupling of nano-indentation and crystal plasticity finite element (CPFE) simulations is widely used to quantitatively probe the small-scale mechanical behaviour of materials. Earlier studies showed that CPFE can successfully reproduce the load-displacement curves and surface morphology for different crystal orientations. Here, we report the orientation dependence of residual lattice strain patterns and dislocation structures in tungsten. For orientations with one or more Burgers vectors close to parallel to the sample surface, dislocation movement and residual lattice strains are confined to long, narrow channels. CPFE is unable to reproduce this behaviour, and our analysis reveals the responsible underlying mechanisms.

Instrumented nano-indentation revolutionised nanoscale hardness testing by enabling continuous and high-sensitivity measurements of load and displacement of the indenter throughout the whole loading and unloading processes [1,2]. As such, even for nanoscale penetration depths, properties such as elastic modulus, hardness, creep parameters, and even fatigue behaviour can be evaluated directly from the load-displacement curve, without the need for imaging of the indent impression [1–6]. The simplicity and usefulness of this analysis have made nano-indentation a widely used tool for characterising bulk or thin-film materials, of crystalline or non-crystalline nature, across both engineering materials and biomaterials [2–4,7]. Particularly, it has transformed the study of mechanical properties of high-energy ion-irradiated bulk materials [8–10], where the damaged layer is only a few microns thick [11–14].

Nano-indentation studies on irradiated materials have shown that irradiation-induced defects cause a significant increase in hardness [8–10,14,15] and modified topology around the indent, [8,9,14,15]. However, quantitatively linking the nano-indent behaviour to the underlying deformation mechanisms occurring across multiple scales is challenging. The reason is that the indentation process involves the evolution of the population of irradiation defects at the nano-scale [16] as well as dislocation dynamics at the meso-scale [1,2,17]. This problem has been partially addressed by combining the experiments with meso-scale 3D crystal plasticity finite element (CPFE) simulations of nano-indentation experiments [7,9,14,15,18,19].

For instance, nano-indentation behaviour of [001]-oriented ion-irradiated tungsten has been studied with this methodology. 3D micro-Laue X-ray diffraction [8,9,20] or high-resolution electron backscatter diffraction (HR-EBSD) [9,14] were used to probe the plastic deformation zone and residual lattice strain around indents. The experiments showed increased load and pile-up around indents in the irradiated material and a more localised plastic deformation zone, due to dislocation-channelling induced strain-softening [8]. CPFE formulations that explicitly included strain-softening successfully reproduced the nano-indentation behaviour of irradiation damaged samples [9,14]. A subsequent study, extended to multiple grain orientations, revealed that the irradiation-induced increased pile-up around indents is exclusive to the [001] orientation [15]. This was also successfully captured by CPFE simulations that applied the same deformation mechanism to different crystal orientations [15].

However, the lack of integration of multiple scales limits the capability of such simulations to accurately capture the complete behaviour. For example, the simulations do not capture features such as dislocation channel formation in irradiated materials [9], though the resultant strain-softening was considered. As such, a complete understanding of the material behaviour remains elusive. In some instances, this may lead to important material properties and/or behaviour remaining unexplained. We demonstrate this here, considering pure tungsten as a prototypical BCC metal. For spherical nano-indents in [110] and [111] oriented grains we find long and narrow strain channels that extend over tens of microns. Surprisingly these channels are not predicted by CPFE simulations. The underlying reasons for this are discussed in the light of the mechanisms controlling their formation, morphology and evolution.

High-purity polycrystalline tungsten (> 99.97 wt. %) sheet (1 mm thick) purchased from Plansee was cut into $10 \times 10 \times 1$ mm$^3$ samples and fully recrystallised in vacuum (< $10^{-5}$ mbar) at 1500 °C for 24 h. The surface was mechanically ground and then electropolished to achieve a high quality, defect-free finish (see details in [12,13]). 500 nm deep spherical nano-indents were made in [001], [110] and [111] oriented grains using a MTS NanoXP indenter and a ~5 μm radius diamond tip (continuous stiffness mode with displacement control). One indent was made in each grain, and several grains were indented for each orientation to confirm repeatability.

Electron channelling contrast imaging (ECCI) was performed on a Zeiss Crossbeam 540 SEM equipped with a backscattered electron (BSE) detector (30 kV, 10 nA). Argus imaging and HR-EBSD around nano-indents was carried out using a Zeiss Merlin SEM with a Bruker eFlash detector (30 kV, 15 nA). The normal EBSD set up, with the sample tilted at 70° to the electron beam, was used for both. The

Argus imaging system consistents of three forward-scattered electron detectors receiving signals from different parts of the diffraction pattern [21]. As such, a display of the diode signals using RGB code will produce a colour-coded map that is highly sensitive to small changes in diffraction patterns. The electron diffraction pattern at each point was recorded at a resolution of 800 ×600 pixels for HR-EBSD. The XEBSD code, developed by Britton and Wilkinson [22,23], was used to quantify small distortions of EBSD patterns relative to a nominally strain-free reference point far from the nano-indent [24,25]. Hence the local deviatoric strain tensor and lattice rotation tensor, relative to the reference point, were determined.

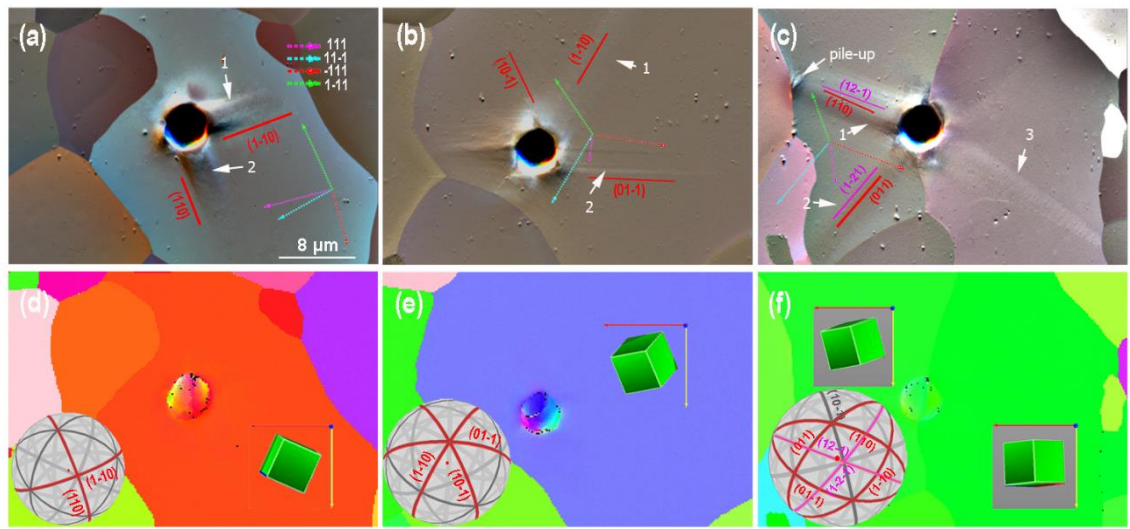

Fig. 1: Orientation dependence of the morphology of nano-indents. (a)-(c) Argus images of the indents in grains closely aligned to <001>, <111> and <110>, respectively. (d)-(f) The inverse pole figure along the out-of-plane direction (IPFZ) of the corresponding Argus images. It should be noted that 'green indent' straddles two <110> grains. The Kikuchi sphere and the unit cube representing the orientation are shown inset. The solid lines in (a-c) represent the traces of planes highlighted in the corresponding Kikuchi spheres. Dash lines represent the <111> directions.

Fig.1 shows the Argus images of the indents in grains closely oriented to <001>, <111> and <110> (a-c). These grains are 12°, 17°, and 14° away from their ideal orientations, respectively. Their inverse pole figures along the out-of-plane (IPFZ) direction are shown in Fig.1 (d-f). Here, we term these three types of grains as red, blue and green according to the colour of the IPF, respectively. Generally, short streaks with pile-up were observed in red grains, but long streaks without pile-up in green and blue grains. These results are reproducible. The long streaks are exclusively visible with Argus, but not in conventional secondary electron image (SEI), due to Argus' high sensitivity to small misorientation [26]. A comparison of SEI and Argus images is shown in supplementary Fig. S1.

In a [001]-oriented single crystal, a pile-up pattern with four-fold symmetry is expected [9]. The structure around the indent in the red grain is slightly asymmetric presumably due to suppression of streaks by the grain boundary (GB). GBs are also relatively close to indents for the other grain orientations, as it is challenging to optically position indents in small grains in well-polished polycrystalline tungsten.

Pile-up streaks in the red grain are ~7 μm long and ~3 μm wide. In the blue grain (Fig. 1(b)), little pile-up is observed around the indent, which agrees well with CPFE predictions [15]. The long streaks in the blue grain (marked as 1 and 2) are up to 16 μm long and ~3 μm wide. The indent in the green grain is located at a GB of two <110> oriented grains (Fig. 1(c)). Here too, streaks up to 20 μm long are observed (labels 2 and 3). Although there is little pile-up near the indent, a large pile-up is seen ~10 μm from the indent, at the intersection of 'streak 1' and a GB (Fig. 1(c)).

In a [001] tungsten single crystal, the pile-up directions align well with the <111> Burgers vector projections (BVPs) on the sample surface [8,9]. Nevertheless, the pile-up directions in the red grain are slightly misaligned with BVPs due to a 14° deviation from ideal [001] orientation (Fig. 1(a)). Closer examination shows that streaks 1 and 2 are parallel to the exit trace of the $(1\bar{1}0)$ and $(110)$ planes, respectively (solid red lines in Fig. 1(a)). No other plane in the {110} or {112} families is aligned to the streaks (see supplementary Fig. S2).

In the blue grain streaks are only observed for two of the BVPs (Fig. 2(b)), though three equivalent BVPs separated by 120° are expected for the ideal [111] orientation. Streak 1 is parallel to $[11\bar{1}]$ with $(1\bar{1}0)$ exit trace, while streak 2 has $(01\bar{1})$ exit trace but is slightly misaligned with the $[\bar{1}11]$ direction (see supplementary Fig. S3). In the left green grain in Fig. 1(c), streaks are found along the $[\bar{1}11]$ and $[11\bar{1}]$ directions. The exit-traces of the $(12\bar{1})$ and $(1\bar{2}1)$ planes are parallel to 'streak 1' and 'streak 2', respectively. The exit-traces of several other planes from the {101} and {12-1} families are also parallel to the streaks (see supplementary Fig. S4).

Table 1 shows the correlation between θ (angle between a Burgers vector and the sample surface) and the streak morphology along the BVPs. A general trend can be observed: Long streaks are seen when the Burgers vector is at a low angle (≤ 20°) to the surface. When all the Burgers vectors are at a large angle to the surface (>20°), pileups or short streaks are observed. This suggests that the surface pile-up depends on the resolved shear strain in the out-of-plane direction.

Residual lattice strains near indents were probed by HR-EBSD (Fig. 2). In the red grain the strain fields are localised close to the indent, corresponding to the short streak and pile-up (Fig. 2(a)). However, in blue and green grains, there are prominent residual lattice strains associated with the long streaks visible in Fig.1(b and c). The long strain channels are also seen in a [101]-oriented tungsten single crystal (Fig. 2(d)). This confirms that the long strain channels are an effect of crystal orientation. For the green grains, the geometric necessary dislocation (GND) density (computed using the approach of [27]) in 'streak 2' and 'streak 3' (labelled in Fig. 2(c)) is higher than the background density. In contrast, the GND density in 'streak 1' is not significantly higher than the background. However, there is a large increase in GND density at the intersection of 'streak 1' and a GB. This means that dislocations had been gliding further away and relatively few are left within the streak.

Table 1: Angle between Burgers vectors and sample surface (θ), and the corresponding pileup and streak morphology observed for different crystal orientations.

| Grain | Burgers vector | [111] | [11$\bar{1}$] | [$\bar{1}$11] | [1$\bar{1}$1] |
|---|---|---|---|---|---|
| Red | θ (°) | 28 | 41 | 45 | 25 |
|  | Morphology | pile-up + short streak or pile-up | | | |
| Blue | θ (°) | 71 | 2 | 20 | 35 |
|  | Morphology | - | long streak | long streak | short streak |
| Green | θ (°) | 59 | 1 | 6 | 50 |
|  | Morphology | - | long streak | long streak | - |

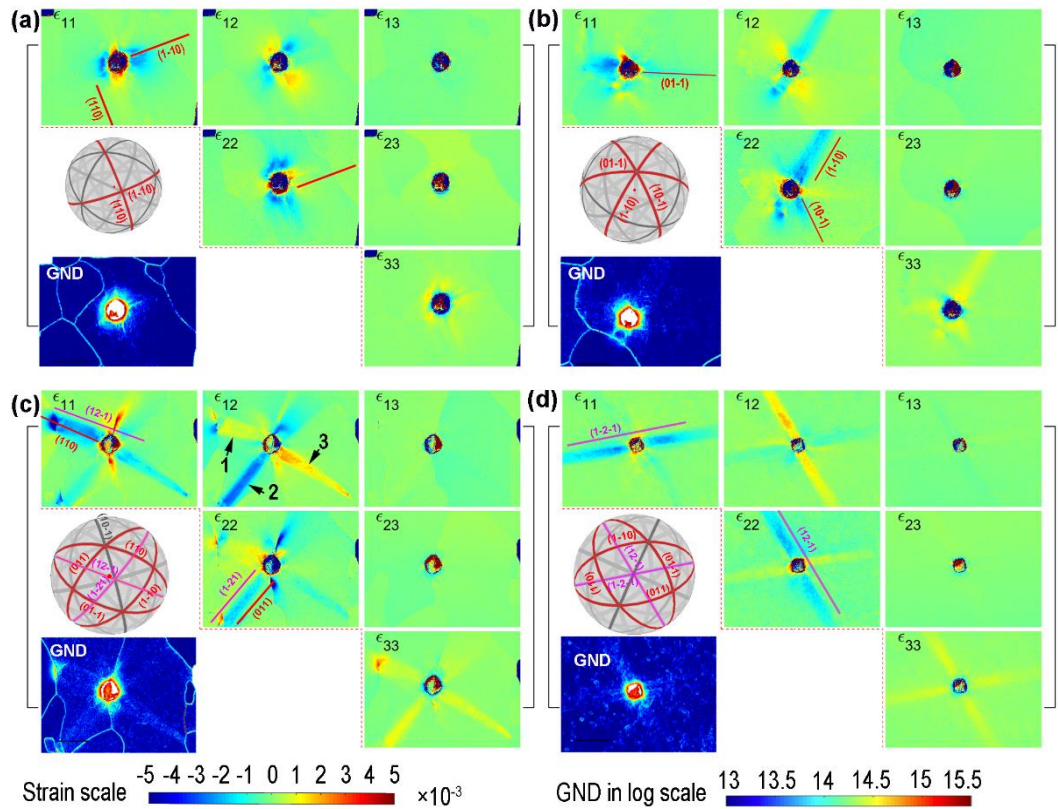

Fig. 2: Maps of 3D deviatoric strain tensor near indents in three different orientations. (a)-(c) correspond to (a - c) in Fig. 1. (d) [101] pure tungsten single crystal.

Fig. 3 (a) and (b) show the dislocation structure in 'streak 3' of the green grain observed by ECCI and GND distribution, respectively. High-density dislocation line segments are only observed in a well-confined channel, which is parallel to the [$\bar{1}$11] direction (θ = 2°) and the exit-traces of the (110), (01$\bar{1}$) and (12$\bar{1}$) planes. The dislocation segments crossing the channel are aligned nearly perpendicular to the channel direction. The fact that these dislocations can glide along with the [$\bar{1}$11] direction for such a long distance within a confined channel suggests that they must have [$\bar{1}$11] Burgers vector. Since the line direction is nearly perpendicular to the Burgers vector, these dislocations must be edge component dominated. These dislocations are presumably gliding on the (101) plane, nearly parallel to the surface, especially if they are prismatic loops. This is in good agreement with a 3D discrete dislocation dynamics (DDD) study of the indentation behaviour in FCC materials [28]. The study showed that prismatic or half prismatic loops form and glide parallel to the surface along the Burgers vector direction when the Burgers vector is near parallel to the surface. However, it cannot be ruled out that some dislocations also glide on the inclined planes such as (110) and (01$\bar{1}$) (see the short segments in Fig. 3(b)) and edge-on (12$\bar{1}$) planes (see dots in Fig. 3(b)).

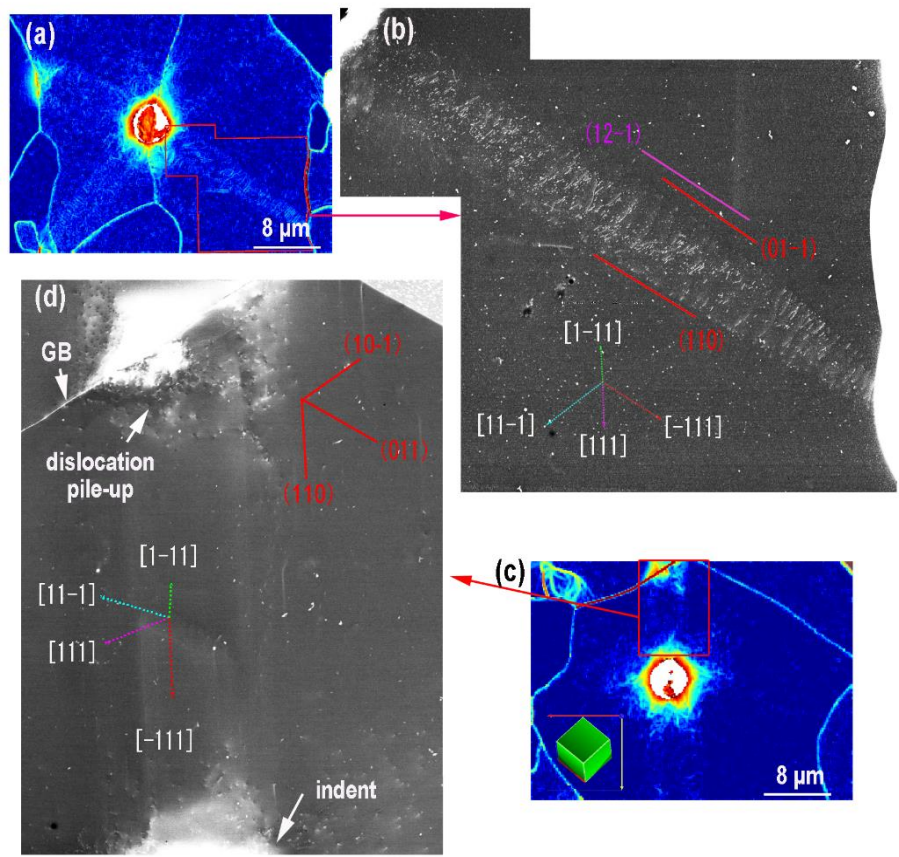

Fig. 3: Dislocation structure within the streaks or dislocation transport channels. (a) and (c) are GND maps, (b) and (d) are corresponding ECCI images of specific streaks.

Fig. 3 (c) and (d) show dislocation structure in a streak of a [1$\bar{1}$3] oriented grain. This is similar to the case of 'streak 1' in Fig. 1(c), where few dislocations are left within the dislocation channel. Rather dislocations pile up at the intersection of the channel with a GB that blocks dislocations' glide (Fig. 3(c)). The pile-up of dislocations at the GB causes an increase of stress that generates plasticity on the other side of the GB. Dislocations in the channel and pile-up show black-white contrast, indicating they are mainly end-on dislocations. The slip traces left by these end-on dislocations are parallel to both the [$\bar{1}$11] (θ = 5°) BVP and the exit-trace of edge-on (110) plane. As such, the dislocations must be of [$\bar{1}$11] type and gliding on the (110) plane.

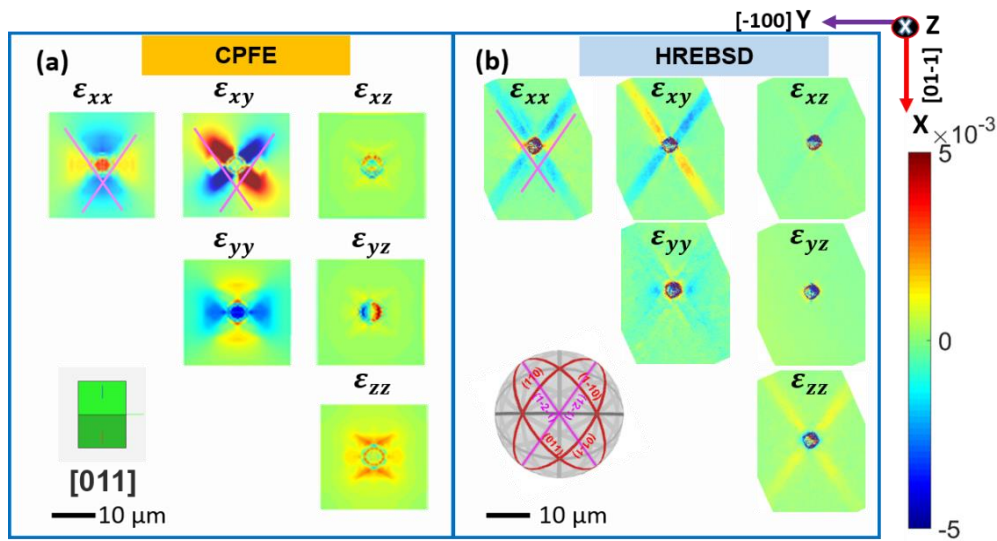

Fig. 4: Comparison of the strain field around an indent in a [011] grain. (a) CPFE simulation. (b) Experimental data from HREBSD of an indent in a [011] single crystal.

The long and confined dislocation transport channels in [110] and [111] orientated grains are not reproduced by CPFE simulations (see Fig. 4 for [110] grain and supplementary Fig. S5 for [111] grain). It should be noted that the same CPFE model successfully reproduced orientation dependence of the surface morphology around indents [15], load-displacement curves [14,15] and the strain field in [100] grains for both un-irradiated and helium ion damaged tungsten [14]. In the CPFE model, a localised strain field around the indent is predicted in contrast to the confined and long streaks seen in experiments. The orientation of CPFE-predicted strain fields is also different from the dislocation channel directions observed in experiments.

Previous 3D DDD of nano-indentation in FCC metals showed that the morphology of dislocation loops punched out by indentation depends on the crystal orientation and pre-existing dislocation sources [28]. In the case of Burgers vector near parallel to the surface, prismatic loops or prismatic half loops, depending on the depth of the dislocation source, can be punched out. These loops thus travel along the

Burgers vector direction [28], transporting material away from the indent. Preliminary small-scale 3D DDD studies on the nano-indentation behaviour of BCC iron suggested a similar behaviour [29]. From a single dislocation source, half loops were formed close to the surface. They preferentially propagate along the Burgers vector direction rather than the lateral direction, along which propagation requires dislocation climb. Therefore, a long streak is expected after indentation when at least one of the Burgers vectors is near parallel to the surface, as observed in this study.

The CPFE constitutive law considers crystallographic slip on the <111>{110} system. However, the nucleation of dislocations, the morphology of dislocation loops and the transport of dislocations are not included. That is, slip can be activated anywhere, even if there is a lack of sources, provided the resolved shear stress exceeds the critical resolved shear stress of a slip system. Furthermore, the effect of dislocation structure and organisation, such as dislocation pile-up, on the stress field is not evaluated in the CPFE model. However, the stress-induced by piling-up of dislocations (Fig. 3(b)) is the main stress driving dislocation motion outside the immediate vicinity of the indent.

In summary, the surface deformation, i.e. surface pile-up is dependent on the resolved magnitude of the Burgers vector along the out-of-plane direction. Surface pile-up forms for Burgers vectors inclined to the surface at an angle greater than ~20°, while long streaks (dislocation transport channels) without pile-up form for Burgers vectors near parallel to the sample surface. CPFE simulations can successfully predict the load-displacement curve and surface morphology for different orientations. However, when it comes to lattice strain, CPFE fails to capture the formation of extended dislocation transport channels. This is because CPFE does not consider the nucleation and morphology of dislocations, or indeed their transport and stress contribution during plastic deformation. Multi-scale simulations are needed to capture these effects in a computationally efficient fashion, offering a coarse-grained representation of the detailed behaviour seen in DDD simulations.

**Acknowledgements**

This work was funded by Leverhulme Trust Research Project Grant RPG-2016-190. E.T. acknowledges financial support from the Engineering and Physical Sciences Research Council Fellowship Grant EP/N007239/1. The authors acknowledge the use of characterisation facilities within the David Cockayne Centre for Electron Microscopy, Department of Materials, University of Oxford and within the LIMA Lab, Department of Engineering Science, University of Oxford.